\documentclass[journal]{IEEEtran}
\usepackage{comment}
\usepackage{cite}
\usepackage{amsmath,amssymb,amsfonts}
\usepackage{algorithmic}
\usepackage{graphicx}
\usepackage{textcomp}
\usepackage{xcolor}
\usepackage{comment}
\usepackage{multirow}
\usepackage{url}
\usepackage{booktabs}
\usepackage{textcomp}
\usepackage{tikz}
\usepackage{subfigure}
\usepackage{array}
\usepackage{siunitx}
\usepackage{booktabs}
\usepackage{tabularx}
\usepackage[normalem]{ulem}

\ifCLASSINFOpdf
\else

\fi

\begin{document}

\title{A Vision Transformer Accelerator ASIC for Sleep Stage Classification in Wearable Devices}

\author{Tristan Robitaille,~\IEEEmembership{Student Member,~IEEE},
        and~Xilin~Liu,~\IEEEmembership{Senior~Member,~IEEE}% <-this % stops a space}
%\thanks{*These authors contributed equally to the work.}

\thanks{This work was supported in part by the Connaught Innovation Award, University of Toronto, the Digital Research Alliance of Canada, and the CMC Microsystems.}

\thanks{Tristan Robitaille was with the Edward S. Rogers Sr. Department of Electrical \& Computer Engineering (ECE), University of Toronto, Toronto, ON, Canada. He is currently with ETH Zurich, Zurich, Switzerland.}
\thanks{X. Liu is with the Edward S. Rogers Sr. Department of Electrical \& Computer Engineering (ECE), University of Toronto, Toronto, ON, Canada. }
}
%}

% The paper headers
\markboth{Submission to IEEE Transactions}%,~Vol.~14, No.~8, August~2015
{Shell \MakeLowercase{\textit{et al.}}: Bare Demo of IEEEtran.cls for IEEE Journals}

% make the title area
\maketitle

\begin{abstract}
This paper presents SleepViT, a custom accelerator ASIC for real-time, low-power sleep stage classification in wearable devices. At the core of SleepViT is a lightweight vision transformer model specifically optimized for electroencephalogram (EEG)-based sleep stage classification. The model is trained on the MASS SS3 dataset and achieves a classification accuracy of 82.9\% across four sleep stages, while requiring only 31.6k weights—demonstrating its suitability for embedded inference. 
The proposed transformer is designed and synthesized in 65nm CMOS technology. To minimize power and area, the architecture adopts a novel layer-dependent fixed-point quantization scheme, variable data widths, and optimized memory access patterns. The synthesized accelerator occupies 0.754\,mm$^2$ of silicon, operates at a maximum clock frequency of 379\,MHz, and consumes 6.54\,mW dynamic and 11.0\,mW leakage power over a 45.6\,ms inference window. With aggressive power gating during idle periods, the effective average power is 0.56\,mW, enabling extended battery life in wearable devices.
This work highlights the feasibility of deploying transformer-based models in highly constrained edge environments and provides a pathway for future biomedical ASICs that require both real-time performance and ultra-low power consumption.
\end{abstract}


\IEEEpeerreviewmaketitle

\section{Introduction}
\label{sec:intro}
Approximately one-third of adults experience insomnia symptoms, while around 8\% suffer from chronic insomnia~\cite{hafner2023societal}.
Chronic insomnia is associated with a wide range of adverse physical and psychological outcomes, including impaired concentration, memory deficits, and reduced attention span, all of which can significantly diminish quality of life and daily productivity.
Acoustic neuromodulation—an emerging, non-invasive technique that employs isochronic tones (IT) to entrain the brain into specific sleep states—has shown promise in enhancing sleep quality, prolonging sleep duration, and offering a potential therapeutic avenue for individuals suffering from insomnia~\cite{kanzler2023effects}. Compared to pharmacological interventions, acoustic neuromodulation carries a lower risk of side effects and dependency, making it a safer alternative for long-term sleep regulation.

To achieve high efficacy, acoustic neuromodulation needs to entrain the brain into the appropriate phase of the sleep cycle through cycle-specific, closed-loop stimulation. This necessitates real-time, in-vivo classification of sleep stages—a process known as \textit{sleep staging}.
The current gold standard for sleep staging relies on polysomnography (PSG), which involves the recording of multiple biosignals such as electroencephalography (EEG), electromyography (EMG), and electrocardiography (ECG). These signals are then manually annotated post-sleep by trained clinicians, a labor-intensive process that typically takes up to two hours per subject~\cite{phan2022automatic}. Accurate sleep staging often requires as many as 19 PSG channels to capture the full spectrum of neural and physiological activity~\cite{RUNDO2019381}. This high degree of instrumentation and manual processing makes PSG unsuitable for continuous, at-home monitoring—especially in the context of a wearable neuromodulation system, where size, power, and user comfort are critical constraints.

% To address the downsides of PSG, machine learning models, initially CNN models, have been proposed for performing sleep staging using a subset of PSG signals \cite{yang2018study, phan2020towards, koushik2018real, chen2020sleep, korkalainen2019accurate, phan2018joint, biswal2017sleepnet}. 
% However, existing approaches either suffer from limited staging accuracy or have models that are too large and computationally intensive, making them unsuitable for deployment in low-power wearable devices for real-time inference~\cite{sun2023design}. \textcolor{red}{Transformers, via their attention mechanism, can capture long-term relationships and learn the temporal characteristics of time-series signals such as the EEG signal used here better than CNNs for a given size budget. This is particularly valuable in the case of sleep staging since the signals of interest evolve over the full length of an inference period. \cite{anwar2024transformers}}

To overcome the limitations of traditional PSG-based approaches, a range of machine learning methods—particularly convolutional neural networks (CNNs)—have been proposed to perform automated sleep staging using a reduced subset of PSG signals~\cite{yang2018study, phan2020towards, koushik2018real, chen2020sleep, korkalainen2019accurate, phan2018joint, biswal2017sleepnet, liu2021system}. While CNN-based models have demonstrated promising results, many either lack sufficient classification accuracy or require large and computationally intensive architectures, rendering them impractical for deployment in resource-constrained, low-power wearable devices designed for real-time operation~\cite{sun2023design}. Recently, transformer-based architectures have gained attention for time-series analysis due to their ability to model long-range temporal dependencies via a self-attention mechanism. Compared to CNNs, transformers can capture global temporal dynamics more effectively within a similar model size budget. This characteristic is particularly beneficial for sleep staging, where signal features often evolve gradually over extended inference windows~\cite{anwar2024transformers}.

In this work, we adopt and extend the Vision Transformer (ViT) architecture~\cite{dosovitskiy2020image} for automated sleep staging using a single-channel EEG as input. While ViTs have demonstrated strong performance across various visual and sequence modeling tasks, their deployment in hardware—particularly for low-power, always-on applications—remains challenging. The self-attention mechanism at the core of ViTs introduces high computational complexity and memory access demands, which are especially taxing in resource-constrained environments. Recent efforts have focused on making ViTs more hardware-friendly through model quantization, pruning, and architecture-aware optimizations~\cite{du2024model,marino2024mevit,liang2023mcuformer}. %For example, ME-ViT proposes a memory-efficient single-load FPGA accelerator for ViTs, targeting embedded applications with limited bandwidth and memory reuse capabilities~\cite{marino2024mevit}. Similarly, MCUFormer explores techniques to fit ViT models into microcontrollers with severe memory limitations~\cite{liang2023mcuformer}. 
However, many of these designs are either evaluated on general-purpose hardware or do not fully meet the stringent power and performance constraints required for wearable biomedical devices.

To bridge this gap, we developed SleepViT—a ViT-based sleep staging model designed as a custom ASIC accelerator. The design is co-optimized for ultra-low power consumption and reliable real-time operation in wearable systems. The ASIC accelerator has been synthesized in TSMC 65nm CMOS technology \textcolor{black}{to evaluate the properties of the solution considering a targeted broad-market wearable sleep tracking device. The 65nm technology node was selected as a trade-off between performance, power, and cost.} With an extra post-softmax rolling window averaging filter, layer-dependent fixed-point format and variable storage data width, the proposed SleepViT accelerator ASIC achieved an accuracy of 82.9\% classifying wake, light, deep and REM sleep stages using an effective power of 0.56mW and an area of 0.754mm$^{2}$. 

\textcolor{black}{
The primary contributions of this work are:
\begin{itemize}
\item Novel integration of a rolling window filter at the softmax output to reduce temporal noise and enhance prediction stability in ViT-based sleep staging;
\item First ASIC implementation of a Vision Transformer for sleep stage classification which is the result of hardware-software co-design combining layer-specific quantization, centralized architecture, and power gating to achieve sub-milliwatt operation while maintaining clinically good accuracy.
\end{itemize}
}

\textcolor{black}{
Key implementation strategies that enable these contributions include:
\begin{itemize}
\item Use of layer-specific fixed-point formats and data widths to optimize memory utilization while maintaining model accuracy;
\item Centralized architecture using a single computing core to eliminate inter-module data movement and silicon area overhead;
\item Instruction-less finite state machine (FSM) control for efficient execution without the need for a microcontroller or instruction decoding; %Instruction-less FSM control to avoid instruction processing overhead and microcontroller involvement;
\item Implementation of all compute functions using hardwired IP blocks with a shared adder and multiplier to minimize silicon area; %Hard IPs for all compute with a shared fundamental adder and multiplier to limit silicon area;
\item Application of power gating at the module level to significantly reduce leakage power. %Power gating at the module level to limit leakage power
\end{itemize}
}

The remainder of this paper is organized as follows: Section~\ref{sec:ViT} introduces the SleepViT model, including the training methodology and performance evaluation. Section~\ref{sec:hardware} presents the design of the custom ASIC accelerator and its simulation results. Section~\ref{sec:discussion} provides a discussion of key findings and design trade-offs, and Section~\ref{sec:conclusion} concludes the paper with a summary and outlook for future work.

\section{SleepViT Model Architecture}
\label{sec:ViT}

\subsection{Model Input and Output}
The proposed model is designed to perform sleep stage classification based on short segments of EEG data, known as \textit{epochs}. Each epoch consists of a 30-s window of 16-bit EEG signals sampled at 128,Hz, resulting in 3,840 data points per input instance. The model is trained to distinguish among four clinically recognized sleep stages: wakefulness, light sleep (N1 and N2), deep sleep (N3), and rapid eye movement (REM) sleep. These stages are defined based on characteristic patterns in EEG frequency and amplitude, and are essential for assessing sleep quality and diagnosing sleep disorders.

\subsection{Model Architecture}
\begin{figure}[!ht]
  \centering
  \includegraphics[width=0.35\textwidth]{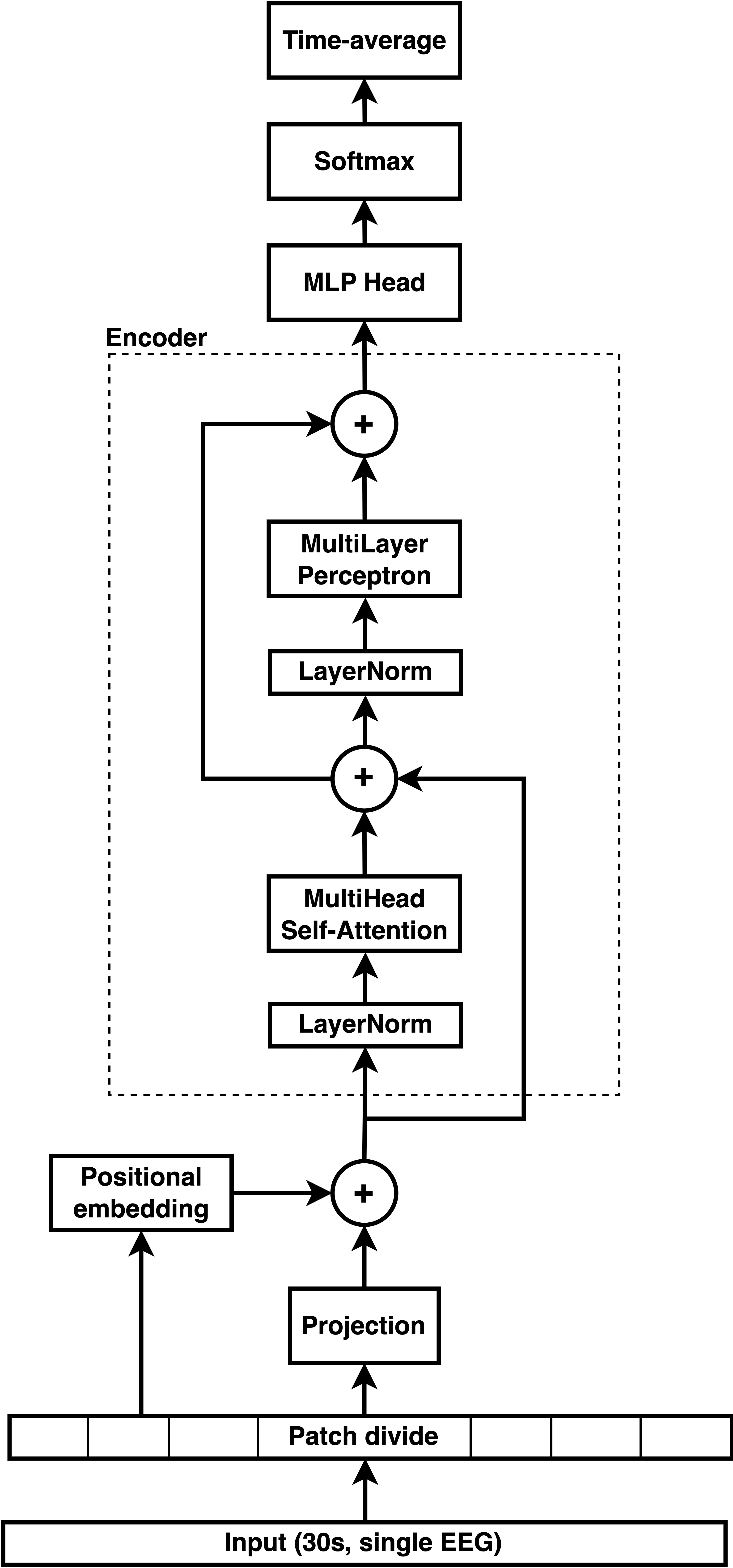}
  \caption{High-level transformer architecture for sleep staging using a single channel EEG}
  \label{fig:vit}
\end{figure}

The ViT, originally introduced by~\cite{dosovitskiy2020image}, is a decoder-less transformer model that uses a sequence of patches as input. In the context of this work, sleep staging is formulated as a ``sequence-to-one’’ classification task, where a 30-second EEG epoch is mapped to a single sleep stage label. As such, there is no requirement for autoregressive feedback or sequential output generation, and the decoder stack used in traditional sequence-to-sequence transformers is unnecessary. The architecture of the proposed model is illustrated in Fig.~\ref{fig:vit}. It consists of a patch embedding layer, a single transformer encoder layer, and a final softmax classifier. To enhance temporal consistency and reduce inference noise, a post-processing step applies a moving average filter to the softmax outputs of the last three predictions before applying the argmax operation. This simple filtering mechanism resulted in a 1.4\% increase in classification accuracy \textcolor{black}{compared to the baseline SleepViT architecture}. To the best of our knowledge, this technique has not been employed in prior AI-based sleep staging models.

%Unlike traditional polysomnography equipment, the target in-ear neuromodulation device is optimized to use a single EEG electrode \cite{voderholzer2012sleep} to maximize the practicality of the overall wearable device. For this reason, the ViT model is designed to use a single EEG electrode.
Unlike traditional PSG systems that rely on multi-channel EEG, EMG, and EOG signals, the target wearable neuromodulation device is designed for practical, at-home use and employs a single EEG electrode~\cite{voderholzer2012sleep}. Accordingly, the ViT model is tailored to operate on data from a single EEG channel, making it well-suited for lightweight, wearable applications.

\subsection{Model Training}
% The TensorFlow model is trained for 100 epochs on the 62 nights of the MASS SS3 dataset using the Cz-LER input channel with a batch size of 16 using the Compute Canada scientific computing cluster \cite{SP3/9MYUCS_2022}. Hyperparameter search and ablation studies are used to maximize accuracy while minimizing model size.

The TensorFlow model is trained for 100 epochs using the Cz–LER EEG channel from the MASS SS3 dataset. Training is performed with a batch size of 16 on the Compute Canada scientific computing infrastructure~\cite{SP3/9MYUCS_2022}. A series of hyperparameter tuning experiments and ablation studies are conducted to optimize the trade-off between model accuracy and architectural efficiency, with a particular focus on minimizing parameter count and memory usage to meet the constraints of wearable hardware deployment.

\subsection{Results \textcolor{black}{and Comparison}}

 % Table~\ref{tab:model_param} lists the model hyperparameters. The 31-fold accuracy of this model on the MASS SS3 data is 82.9\%, and it has only 31.6k weights. Fig.~\ref{fig:confusion_matrix} provides the confusion matrix for the 31-fold accuracy evaluation. As can be seen, REM and light sleep are the most accurately predicted sleep stages. Table~\ref{tab:model_comparison} compares it to state-of-the-art models published in the literature. The vision transformer presented in this work maintains reasonable accuracy while being 8 times smaller than the next smallest model, which makes it ideal for a wearable application. It can maintain adequate accuracy due to the final denoising moving average filter.

\begin{figure}[!ht]
  \centering
  \includegraphics[width=0.32\textwidth]{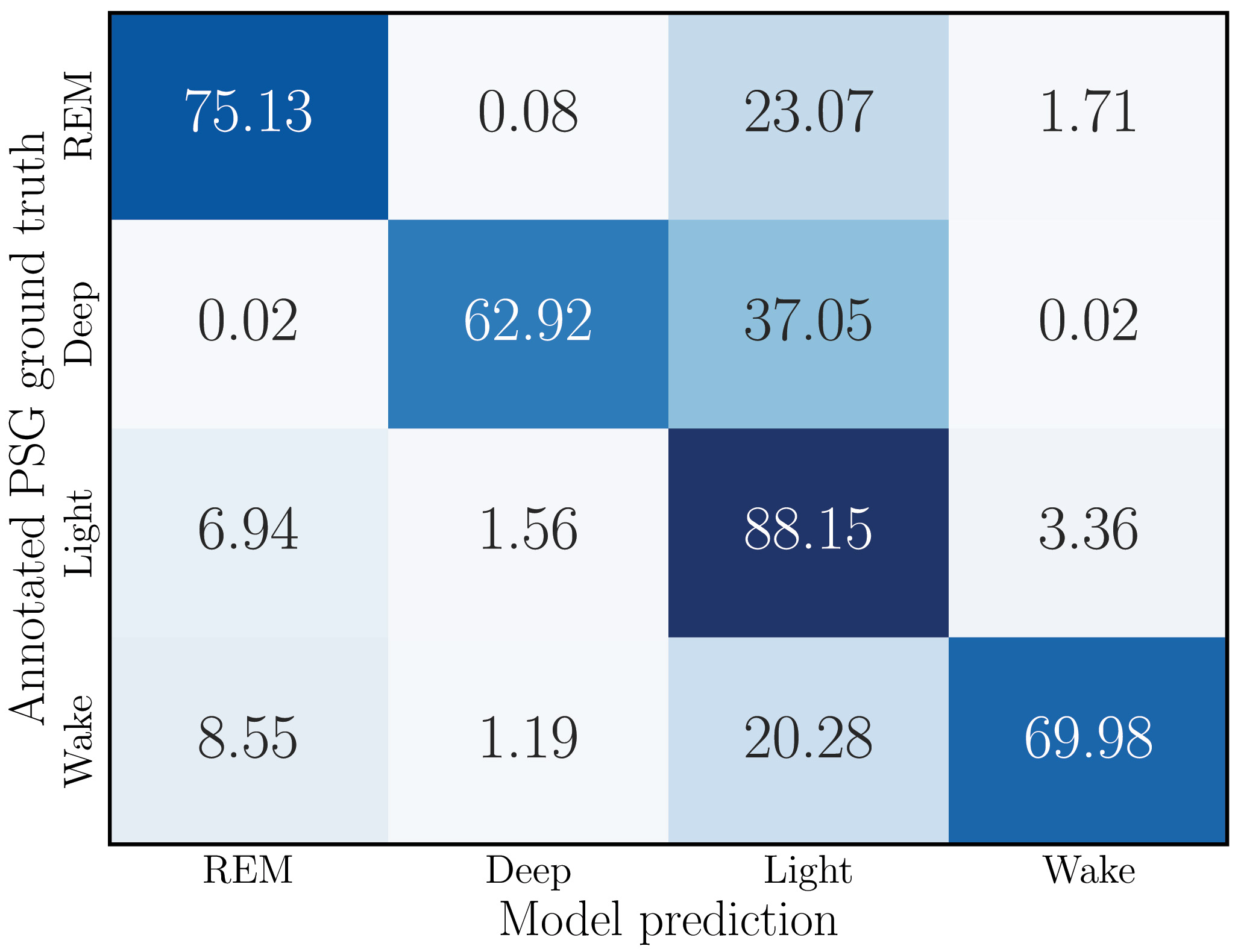}
  \caption{Confusion matrix for the 31-fold accuracy evaluation}
  \label{fig:confusion_matrix}
\end{figure}

The model’s hyperparameters are summarized in Table~\ref{tab:model_param}. Evaluated using 31-fold cross-validation on the MASS SS3 dataset, the model achieves an accuracy of 82.9\% with only 31.6k trainable parameters. The corresponding confusion matrix is shown in Fig.~\ref{fig:confusion_matrix}, illustrating that REM and light sleep stages are classified with the highest accuracy.

\begin{table}[!ht]
  \centering
  \renewcommand{\arraystretch}{1.2}
  \setlength{\arrayrulewidth}{1.5pt}
  \caption{Hyperparameters for the vision transformer model}
  \begin{tabularx}{0.4\textwidth}{@{} *1X *1l @{}}
      \toprule
      Metric                          & Value \\\midrule
      Input channel                   & Cz-LER        \\
      Size (\# of weights)            & 31,589        \\
      Sampling frequency              & 128 Hz        \\
      Clip length                     & 30s           \\
      Patch length                    & 64 samples    \\
      Embedding depth ($d_{model}$)   & 64            \\
      \# of attention heads           & 8             \\
      \# of encoder layers            & 1             \\
      MLP dimension                   & 32            \\
      MLP head depth                  & 1             \\
      Output averaging depth          & 3 samples     \\ \bottomrule 
  \end{tabularx}
  \label{tab:model_param}
\end{table}

% A comparative analysis against state-of-the-art models is presented in Table~\ref{tab:model_comparison}. The proposed SleepViT model achieves competitive accuracy while being approximately 8x smaller than the next most compact model reported in the literature. This substantial reduction in model size, combined with its strong performance, makes it particularly well-suited for deployment in resource-constrained wearable systems. The use of a post-softmax moving average filter plays a key role in preserving accuracy by mitigating prediction noise in real-time inference.

Table~\ref{tab:model_comparison} provides a comparative analysis with existing state-of-the-art models from the literature, regardless of whether they were implemented on hardware. Despite its compact architecture, the proposed SleepViT model maintains competitive accuracy and is approximately \textcolor{black}{5.7x} smaller than the next most compact baseline. This significant reduction in model size, combined with its ability to run efficiently on low-power hardware, underscores its suitability for deployment in wearable systems. The inclusion of a lightweight, post-softmax moving average filter contributes to this performance by smoothing transient fluctuations in predictions, resulting in a 1.4\% gain in classification accuracy \textcolor{black}{compared to the baseline SleepViT architecture}. {\color{black} Note that some low-power models, such as BiLSTM-based networks, have been proposed in the literature and have achieved compact model sizes, for example, 248k parameters in ZleepAnlystNet~\cite{zleepanlystnet}. However, we did not include them in the comparison table, as they were evaluated on a different dataset.}

\begin{table*}[!ht]
  \centering
  \renewcommand{\arraystretch}{1.2}
  \setlength{\arrayrulewidth}{1.5pt}
  \caption{Comparison with state-of-the-art sleep staging models {\color{black}evaluated on MASS-SS3}}
  \begin{tabular}{@{} >{\arraybackslash}p{.8cm} >{\centering\arraybackslash}p{3.3cm} >{\centering\arraybackslash}p{3cm} >{\centering\arraybackslash}p{2cm} >{\centering\arraybackslash}p{2cm} >{\centering\arraybackslash}p{1.25cm} >{\centering\arraybackslash}p{2.25cm} @{}}
    \toprule
    Year                   & Work                                                             & Model                               & Input                         & Accuracy                           & Size                   & Hardware Implementation \\\midrule
    2017                   & DeepSleepNet \cite{supratak2017deepsleepnet}                     & CNN/LSTM                            & Fpz-Cz                        & 86.2\%                             & 24.7M                  & \multirow{8}{*}{\textcolor{black}{No}} \\
    \textcolor{black}{2018} & \textcolor{black}{Chambon \textit{et al.} \cite{chambon2018deep}} & \textcolor{black}{CNN}               & \textcolor{black}{2 channels}  & \textcolor{black}{78\%}             & \textcolor{black}{N/A}  &  \\
    \textcolor{black}{2018} & \textcolor{black}{Phan \textit{et al.} \cite{phan2018joint}}      & \textcolor{black}{CNN}               & \textcolor{black}{C4-A1}       & \textcolor{black}{78.6\%}           & \textcolor{black}{N/A}  &  \\
    \textcolor{black}{2019} & \textcolor{black}{IITNet \cite{seo2020intra}}                     & \textcolor{black}{CNN/RNN}           & \textcolor{black}{F4-EOG}      & \textcolor{black}{86.6\%}           & \textcolor{black}{N/A}  &  \\
    \textcolor{black}{2020} & \textcolor{black}{TinySleepNet \cite{supratak2020tinysleepnet}}   & \textcolor{black}{CNN/RNN}           & \textcolor{black}{F4-EOG}      & \textcolor{black}{\textbf{87.5\%}}  & \textcolor{black}{1.3M} &  \\
    \textcolor{black}{2021} & \textcolor{black}{EOGNET \cite{fan2021eognet}}                    & \textcolor{black}{CNN/RNN}           & \textcolor{black}{EOG}         & \textcolor{black}{83.1\%}           & \textcolor{black}{1.7M} &  \\
    \textcolor{black}{2021} & \textcolor{black}{RobustSleepNet \cite{guillot2021robustsleepnet}}& \textcolor{black}{BiGRU + Attention} & \textcolor{black}{10 channels} & \textcolor{black}{82.2/84.0/80.8\%} & \textcolor{black}{180k} &  \\
    \textcolor{black}{2024} & \textcolor{black}{SPDTransNet \cite{seraphim2024structure}}       & \textcolor{black}{Transformer}       & \textcolor{black}{8 channels}  & \textcolor{black}{84.9/84.4/84.3\%} & \textcolor{black}{N/A}  &  \\
    \bottomrule
    2025 & This work                                                                        & Vision Transformer                    & Cz-LER                        & 82.9\%                             & \textbf{31.6k}         & Yes \\
    \hline
  \end{tabular}
  \label{tab:model_comparison}
\end{table*}

\section{ASIC Implementation of the SleepViT Accelerator}
\label{sec:hardware}
The model presented in Section \ref{sec:ViT} has been implemented as a custom ASIC to meet the stringent energy efficiency requirements of wearable applications. In this work, we report synthesis results as preliminary estimates of the accelerator’s power consumption and silicon area. The hardware architecture is described in SystemVerilog, functionally verified using Verilator, synthesized using Synopsys Design Compiler, and physically implemented with Cadence Innovus for layout generation.

\begin{figure}[!ht]
  \centering
  \includegraphics[width=0.45\textwidth]{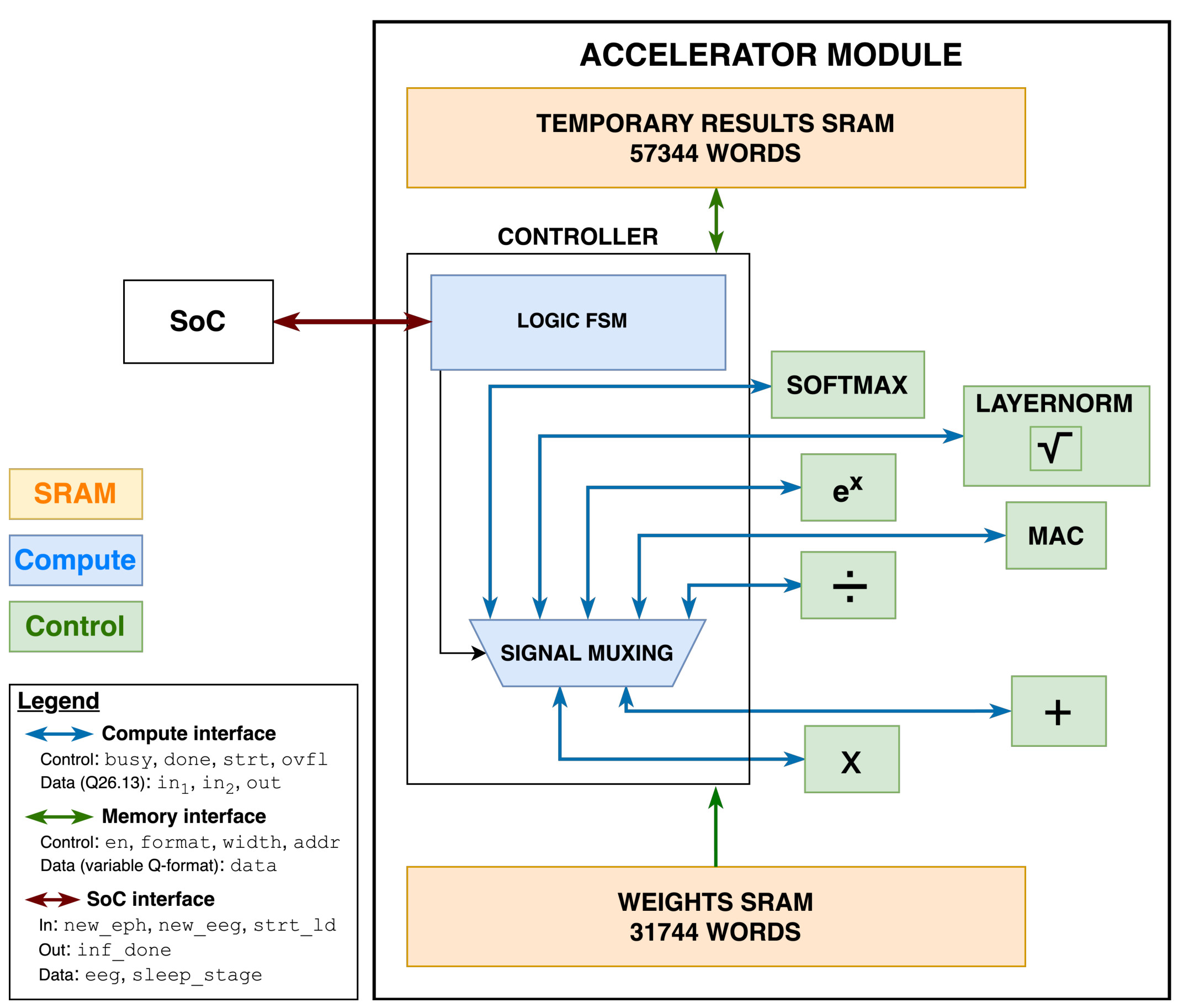}
  \caption{High-level vision transformer accelerator architecture}
  \label{fig:arch}
\end{figure}

\begin{figure*}[!ht]
  \centering
  \includegraphics[width=.95\textwidth]{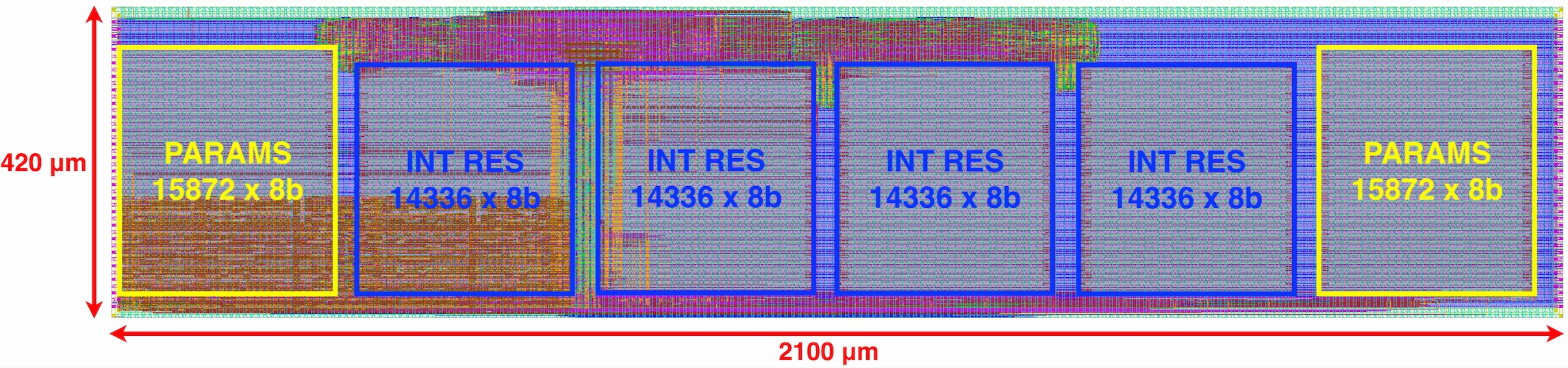}
  \caption{Synthesized layout of the SleepViT ASIC accelerator in 65nm CMOS}
  \label{fig:die_shot}
\end{figure*}

Fig.~\ref{fig:die_shot} presents the layout generated by Cadence Innovus, which occupies a total silicon area of 2,100\si{\micro\meter} by 420\si{\micro\meter}. As shown, the on-chip memory takes a significant portion of the silicon area.

\subsection{Hardware Architecture}
\begin{figure}[!ht]
          \centering
  \includegraphics[width=0.45\textwidth]{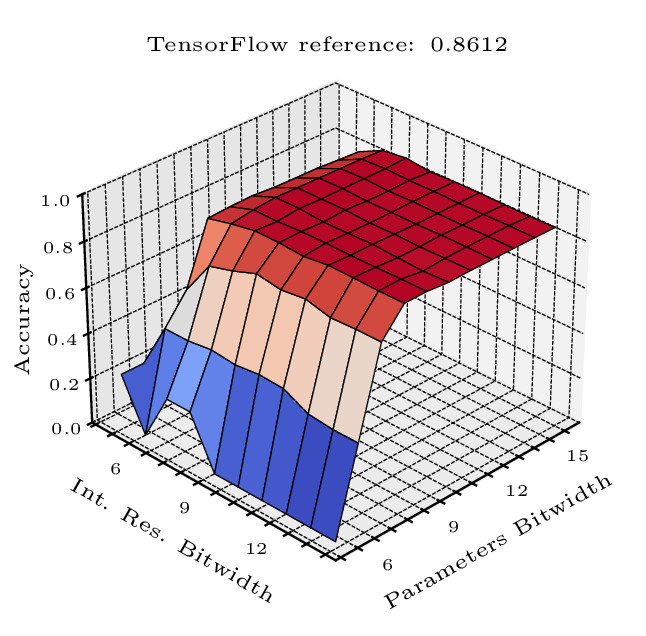}
  \caption{Inference accuracy as a function of data width}
  \label{fig:fix_pt_acc_study}
\end{figure}

Fig.~\ref{fig:arch} shows the high-level architecture of the ASIC accelerator module. The accelerator instantiates its own SRAM for storing weights and intermediate results. To minimize area, leakage power, and communication overhead, the accelerator features a single compute core composed of a control finite state machine (FSM) and multiple dedicated hard IP blocks for core computational tasks. The accelerator is capable of performing inference autonomously, without requiring any intervention from an external host processor. The inference control logic is implemented as an instruction-less FSM, eliminating the need for firmware or microcontroller support. Input EEG data is provided as 16-bit unsigned integers.

To optimize resource utilization, fundamental arithmetic blocks—including adders and multipliers—are shared across all computation modules and the FSM. The FSM orchestrates operation sequencing by multiplexing data paths and control signals between the hard IP blocks. For compatibility with the rest of the system, the clock frequency is fixed at 100\si{\mega\hertz} in simulations.

\subsection{Memory}
The accelerator uses a layer-dependent, single- (8-bit) or double-width (16-bit) fixed-point format for storage in local SRAM. This allows for maximum dynamic range over the inference path while maintaining a low bitwidth. Fig.~\ref{fig:fix_pt_acc_study} presents the accuracy of a sample night as a function of the number of bits used for weights and intermediate results memory. On this sample night, the reference accuracy is 86.12\% and the accuracy of the model with 8-bit weights and 8-bit intermediate results is 86.01\%. This data is sourced from a C++ functional simulation of the hardware architecture which uses Xilinx’s HLS library for fixed-point arithmetic \cite{xilinxhls}. Weights are always single-width and, depending on the layer of the ViT, are stored in Q2.6 to Q5.3 formats. Intermediate results are stored in Q1.7 to Q6.2 for single-width data or Q8.8 for double-width data. The memory for weights is split into two identical banks (each 15872 words), and the memory for intermediate results is split into 4 identical banks (each 14336 words). The SRAM macro is compiled with ARM Artisan IP. To permit single-cycle latency for double-width data, each half is stored in a different bank and both banks are accessed simultaneously by the memory controller modules. The memory controllers also handle casting the data to and from the appropriate fixed-point format as requested by the ViT FSM controller.

\subsection{Computing Modules}
This section outlines the design details of the key computational modules, including the adder, multiplier, divider, exponential unit, square-root unit, multiply-and-accumulate (MAC) module, softmax, and layer normalization modules.

\subsubsection{Adder and Multiplier}
% The adder and multiplier are combinational modules that add and multiply, respectively, two fixed-point numbers. The adder uses a ripple-carry adder architecture. They both have a single-cycle latency, which simplifies the logic that uses it. They both provide an \texttt{overflow} flag. To reduce dynamic power consumption, the outputs are only updated when the \texttt{refresh} signal is high. Both modules implement symmetrical saturation and truncation towards \(-\infty\). Note that the adder and multiplier modules are shared by all higher-level modules through a MUX implemented in the transformer controller.

The adder and multiplier are implemented as combinational modules that operate on two fixed-point inputs. The adder utilizes a ripple-carry architecture, while both modules are designed with single-cycle latency to simplify downstream control logic and minimize pipeline complexity. Each module includes an \texttt{overflow} flag for error detection during arithmetic operations.

To reduce dynamic power consumption, outputs are only updated when the \texttt{refresh} signal is asserted. Both units support symmetrical saturation and truncation toward \(-\infty\), ensuring consistent handling of fixed-point edge cases. Importantly, the adder and multiplier are shared resources across all higher-level functional blocks. Access to these arithmetic units is managed via a multiplexer within the transformer controller, enabling efficient reuse without area duplication.

\subsubsection{Divider}
\label{sec:divider_desc}
The divider is more complex than the adder and multiplier. It performs bit-wise long-division and has a latency of \(N+Q+3\) cycles, where \(N\) is the total number of bits of the fixed-point format and \(Q\) is the number of fractional bits. %The divider also provides flags for \texttt{overflow} and \texttt{divide-by-zero} and \texttt{done}/\texttt{busy} status signals. The divider module implements Gaussian rounding. Like all other compute modules below, the dividers starts division on an active-high pulse of the \texttt{start} signal and pulses the \texttt{done} signal when the result computation is finished.
The divider provides additional control and status signals, including \texttt{overflow}, \texttt{divide-by-zero}, \texttt{done}, and \texttt{busy} flags for robust integration into the control flow. Gaussian rounding is used to improve numerical accuracy in fixed-point representation. As with other computational modules, division is initiated by a rising edge on the \texttt{start} signal, and the module asserts the \texttt{done} signal upon completion of the operation.

\subsubsection{Exponential Unit}
\begin{figure}[!ht]
  \centering
  \includegraphics[width=0.45\textwidth]{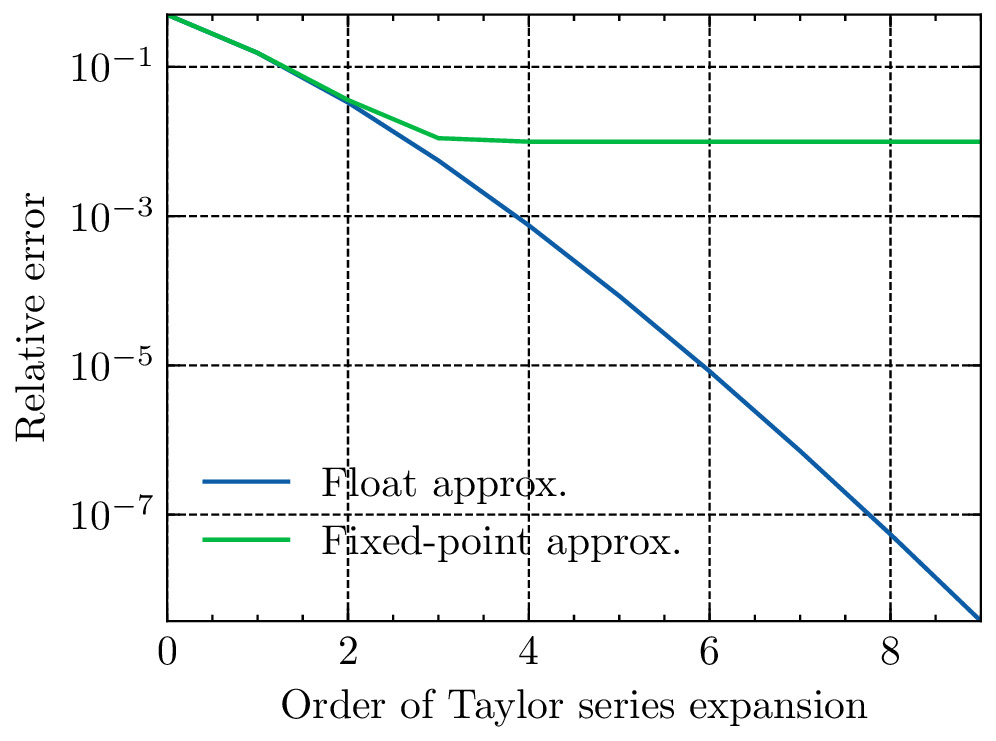}
  \caption{Error of the exponential function approximation for fixed-point and float formats}
  \label{fig:exp_accuracy}
\end{figure}

The exponential unit computes the natural exponential, \(e^x\), of a fixed-point number \(x\). It uses a combination of the identities of exponentials and a Taylor series approximation around zero to compute the exponential. Specifically, the module transforms the
exponential as such:
\[
e^{x} = 2^{\frac{x}{\ln(e)}} = 2^{z} = 2^{\lfloor z \rfloor} 2^{z-\lfloor z \rfloor}
\]
The module computes \(2^{\lfloor z \rfloor}\) an inexpensive bit-shift operation and \(2^{z-\lfloor z \rfloor}\) as a Taylor series approximation with pre-computed coefficients. Figure \ref{fig:exp_accuracy} shows the accuracy of the approximation as a function of the number of terms in the Taylor expression. After 3 terms, the error is dominated by the fixed-point format rather than the approximation, so only the first 3 terms are used in this work. The latency of the exponential module is 24 cycles.

\subsubsection{Square-root Unit}
This modules uses a digit-by-digit algorithm to compute the square root of a fixed-point number. In addition to the control signals described in section \ref{sec:divider_desc}, the square root module also notifies the controller of negative radicand. The latency is \(\lfloor\frac{N+Q}{2}\rfloor+1\).

\subsubsection{Multiply-and-Accumulate (MAC) Module}
The MAC module performs a vector dot-product for a given pair of base addresses for the data and length of the vector and applies a selectable activation function (\texttt{none}, \texttt{linear} or \texttt{swish}) to the result. As reported by Ramachandran \textit{et al}, \texttt{Swish} is similar to \texttt{SiLU} and helps resolve the vanishing gradient problem in backpropagation, leading to a higher accuracy for a given number of training epochs \cite{ramachandran2017searching}. For a nominal length of 64 (which corresponds to the embedding depth of the model, a very common value for matrix dimensions in the model) and Q18.21 format, the latencies are 72, 76 and 170 for the \texttt{none}, \texttt{linear} and \texttt{swish} activation functions, respectively. The multiplication and addition operations are pipelined to maximize throughput. Note that, although the \texttt{swish} activation function comprises a divider operation, the MAC compute latency can still be kept fairly short because the divisor is the same for all elements. The module can thus perform the division once and multiply by the inverse, which is a single-cycle operation. Finally, the MAC module can be directed to choose the second vector from weights or intermediate results memory as needed by different computation in the network.

\subsubsection{Softmax Module}
The softmax module computes the softmax function of a vector of fixed-point numbers:
\[
\sigma(\mathbf{z})_i = \frac{e^{z_i}}{\sum_{j=1}^{K} e^{z_j}}
\]
% For a 64-element Q18.21 vector, the latency is 1926 cycles. This is significantly longer than other vector compute modules such as the MAC because, in the softmax operation, each element is exponantiated individually. The softmax uses the same control signals as the other modules.
For a 64-element input vector in Q18.21 fixed-point format, the softmax module exhibits a latency of 1,926 cycles. This is significantly higher than other vector-based compute modules, such as the MAC module, due to the element-wise exponential operations required by the softmax function. Each input element needs to be individually exponentiated and then normalized across the vector, contributing to the increased computational cost. The softmax module adheres to the same control protocol as the other compute blocks, including the use of standard \texttt{start} and \texttt{done} signaling.

\subsubsection{LayerNorm Module}
% The final compute module is the LayerNorm module. It computes the Layer Normalization of a vector of fixed-point numbers. The LayerNorm operation consists of a normalization of the vector on the horizontal dimension followed by scaling and shifting using learnable parameters on the vertical dimension. For a 64-element Q18.21 vector, the latency is 1943 cycles.

The final computational block in the accelerator is the LayerNorm module, which performs layer normalization on a vector of fixed-point inputs. The operation first normalizes the vector along the horizontal dimension by computing the mean and variance, followed by scaling and shifting along the vertical dimension using learnable parameters. This two-step process improves training stability and convergence, particularly in transformer-based architectures. For a 64-element input vector in Q18.21 fixed-point format, the module exhibits a latency of 1,943 cycles. The implementation is fully pipelined.

\subsection{Power Gating}
Since inference for sleep stage is only required once every 30-\si{\second}, a significant portion of the accelerator’s energy consumption arises from leakage current during idle periods. This idle power can be significantly reduced through power gating, which disables unused modules to suppress leakage. Sathanur \textit{et al.} report that power gating can reduce leakage current by up to 95\% in 65nm CMOS technology—the same process node used in this work~\cite{sathanur2008quantifying}.

Although the PDK used in this work for simulation does not offer power gating cells, we estimate the potential benefit of power gating through analysis of the trace while running a full inference. We augment the testbench to report the ratio of time for which a compute module is active, as shown in Table~\ref{tab:percent_active}. 

\begin{table}[!ht]
  \centering
  \renewcommand{\arraystretch}{1.2}
  \setlength{\arrayrulewidth}{1.5pt}
  \caption{Active ratio over inference and inference period}
  \begin{tabularx}{0.35\textwidth}{@{} *1l >{\centering\arraybackslash}p{2cm} >{\centering\arraybackslash}p{2cm} @{}}
      \toprule
      Module      & Active ratio (over inference) & Active ratio (over 30\si{\second})\\\midrule
      Adder       & 64.1\% & 0.09756\%\\
      Multiplier  & 72.4\% & 0.11015\%\\
      Divider     & 3.53\% & 0.00054\%\\
      Exponential & 16.7\% & 0.02541\%\\
      Square Root & 0.081\%& 0.00012\%\\
      MAC         & 66.9\% & 0.10177\%\\
      Softmax     & 19.7\% & 0.02992\%\\
      LayerNorm   & 5.986\%& 0.00091\%\\
      Memory      & 100\%  & 0.15210\%\\
      \bottomrule
    \end{tabularx}
  \label{tab:percent_active}
\end{table}

\begin{table*}[!ht]
  \centering
  \renewcommand{\arraystretch}{1.2}
  \setlength{\arrayrulewidth}{1.5pt}
  \caption{Performance metrics of the vision transformer accelerator}
  \begin{tabular}{@{} >{\arraybackslash}p{2cm} >{\centering\arraybackslash}p{2cm} >{\centering\arraybackslash}p{2.5cm} >{\centering\arraybackslash}p{2cm} >{\centering\arraybackslash}p{2.2cm} >{\centering\arraybackslash}p{1.75cm} @{}}
    \toprule
    Module        & Area                          & Latency (cycle/op)      & Dynamic power         & Leakage power         & $F_{max}$ \\\midrule
    Adder         & 914.4\si{\mu\square\meter}    & 1                       & 9.83\si{\mu\watt}     & 23.7\si{\mu\watt}     & 3.33\si{\giga\hertz} \\
    Multiplier    & 9914.8\si{\mu\square\meter}   & 1                       & 50.1\si{\mu\watt}     & 245\si{\mu\watt}      & 2.38\si{\giga\hertz} \\
    Divider       & 3005.6\si{\mu\square\meter}   & 63                      & 9.1\si{\mu\watt}      & 50.1\si{\mu\watt}     & 0.521\si{\giga\hertz} \\
    Exponential   & 4071.6\si{\mu\square\meter}   & 24                      & 20.4\si{\mu\watt}     & 75.3\si{\mu\watt}     & 0.862\si{\giga\hertz} \\
    Square Root   & 2270.9\si{\mu\square\meter}   & 31                      & 9.6\si{\mu\watt}      & 41.8\si{\mu\watt}     & 0.379\si{\giga\hertz} \\
    MAC           &                               & 72                      &                       &                       &                     \\
    MAC - linear  & 5656.3\si{\mu\square\meter}   & 76                      & 75.4\si{\mu\watt}     & 113\si{\mu\watt}      & 0.689\si{\giga\hertz} \\
    MAC - swish   &                               & 170                     &                       &                       &                     \\
    Softmax       & 28420.2\si{\mu\square\meter}  & 1926                    & 106.2\si{\mu\watt}    & 562\si{\mu\watt}      & 0.952\si{\giga\hertz} \\
    LayerNorm     & 5055.1\si{\mu\square\meter}   & 1943                    & 51.6\si{\mu\watt}     & 97.9\si{\mu\watt}     & 0.885\si{\giga\hertz} \\
    Weights       & 0.243\si{\milli\square\meter} & 1                       & 18.8\si{\mu\watt}     & 2.75\si{\milli\watt}  & 3.33\si{\giga\hertz} \\
    Int. res.     & 0.442\si{\milli\square\meter} & 1                       & 5.79\si{\milli\watt}  & 6.87\si{\milli\watt}  & 1.75\si{\giga\hertz} \\
    \bottomrule
    Total         & 0.754\si{\milli\square\meter} & 45.6\si{\milli\second} @ 100\si{\mega\hertz}    & 6.54\si{\milli\watt}  & 11.0\si{\milli\watt}  & 0.379\si{\giga\hertz} \\
    \hline
  \end{tabular}
  \label{tab:compute_modules}
\end{table*}

Both memories are left active for the duration of the inference, but power gated between inferences. Based on this activity profiling and assuming that each compute module is power gated when idle—with a 95\% reduction in leakage during gated periods—the effective average power consumption of the accelerator over a 30-s window is estimated to be 0.56\si{\milli\watt}.

\subsection{\textcolor{black}{Results and Comparison}}
Table~\ref{tab:compute_modules} lists the key metrics of each computing blocks and for the overall design over one inference. Inference takes 45.6\si{\milli\second} and can be computed with 0.754\si{\milli\square\meter} of silicon. A processing latency of 45.6\si{\milli\second} represents only 0.15\% of one sleep staging epoch, which amounts to negligible phase offset between measurement and inference availability. Power consumption during inference is 17.5\si{\milli\watt}, with a total leakage power of 11.0\si{\milli\watt}. The maximum frequency of the design, determined by the square root module, is 379\si{\mega\hertz}. LayerNorm is the slowest module, with latency measured as the combined time for normalization as well as centering and scaling.
%Fig.~\ref{fig:die_shot} presents the PnR layout generated by Cadence Innovus. As can be seen, memory consumes the vast majority of the area.
Finally, Table~\ref{tab:summary} presents a consolidated summary of the key performance, power, and area metrics achieved in this work.

\begin{table}[!ht]
    \centering
    \renewcommand{\arraystretch}{1.2} % Vertical spacing
    \setlength{\arrayrulewidth}{1.5pt} % Thickness of vertical lines
    \caption{Summarized results of the model and hardware design}
    \begin{tabularx}{0.35\textwidth}{Xlc}
        \toprule
        Metric                      & Value                        \\\midrule
        31-fold accuracy            & 82.9\%                       \\
        Size                        & 31.589\si{\kilo\byte}        \\ \bottomrule 
        %%%%%%%%%%%%%%%%%%%%%%
        Inference latency           & 45.6\si{\milli\second}       \\
        Area                        & 0.754\si{\square\milli\meter}\\
        Leakage power               & 6.54\si{\milli\watt}         \\
        Average dynamic power       & 11.0\si{\milli\watt}         \\
        Average total power         & 0.56\si{\milli\watt}       \\
        $f_{Max}$                   & 379\si{\mega\hertz}          \\ \bottomrule
    \end{tabularx}
    \label{tab:summary}
\end{table}

\begin{table*}[!ht]
  \centering
  \renewcommand{\arraystretch}{1.2}
  \setlength{\arrayrulewidth}{1.5pt}
  \caption{Comparison with existing ASIC-based sleep staging accelerators}
  \begin{tabular}{@{} 
        >{\arraybackslash}p{1.8cm}
        >{\centering\arraybackslash}p{1.25cm}
        >{\centering\arraybackslash}p{2.2cm}
        >{\centering\arraybackslash}p{1.8cm}
        >{\centering\arraybackslash}p{1cm}
        >{\centering\arraybackslash}p{1cm}
        >{\centering\arraybackslash}p{0.75cm}
        >{\centering\arraybackslash}p{0.75cm}
        >{\centering\arraybackslash}p{1.25cm} @{}}
    \toprule
    Year \& Journal & Work                                                  & Model/Algorithm            & Input         & Accuracy & Area                          & Node                & Power                & Dataset \\\midrule
    2019 TCAS-I     &  Chang \textit{et al.} \cite{chang2019ultra}          & WPD and NN decision tree   & EEG/EEG+EMG & 81.1\%   & 11.72\si{\milli\square\meter}   & 180\si{\nano\meter} & 4.96\si{\micro\watt} & NSRR, NTHU   \\
    2017 JSSC       &  Imtiaz \textit{et al.} \cite{imtiaz2017ultralow}     & Decision tree              & EEG           & 78.9\%   & 10.3\si{\milli\square\meter}  & 180\si{\nano\meter} & 0.58\si{\milli\watt} & MASS         \\
    2016 TBioCAS    &  Kassiri \textit{et al.} \cite{kassiri2016electronic} & Filtering and thresholding & 2 EEG + EMG   & 79.7\%   & -                             & 130\si{\nano\meter} & 0.35\si{\milli\watt} & 9 mice       \\
    \bottomrule
    This work       &                                                       & ViT                        & EEG           & 82.9\%   & 0.754\si{\milli\square\meter} & 65\si{\nano\meter}  & 0.56\si{\milli\watt} & MASS         \\
    \hline
  \end{tabular}
  \label{tab:hw_comparison}
\end{table*}

\textcolor{black}{
Several prior studies have explored hardware implementations for low-power sleep staging. Table~\ref{tab:hw_comparison} summarizes and compares the key characteristics of these efforts. Notably, none of the existing works utilize transformer-based architectures, which are well-suited for sequence modeling and have recently become state-of-the-art in deep learning.
}

% \subsection{Comparison with other hardware accelerators}
% Other authors have designed hardware with the purpose of low-power sleep staging. Table \ref{tab:hw_comparison} compares the key characteristics of each work. No previous work implements a vision transformer for a sleep staging wearable. Rather, Chang \textit{et al.} designed a manual algorithm based on a wavelet packet decomposition (WPD) pre-processing tree followed by extraction of 23 features with neural network (NN)-based decision tree classifier \cite{chang2019ultra}. Their design supports two mode: EEG + EMG or EEG-only input. Implemented in 180nm, the design measures 11.72\si{\milli\square\meter} and consumes 4.96\si{\micro\watt}. The accuracy is 81.1\% on the datasets from NSRR and NTHU \cite{dean2016scaling}, although it is not measured with the standard k-fold validation but rather measured using true and false positives on a subset of samples for each sleep stage.

\textcolor{black}{
Chang \textit{et al.}~\cite{chang2019ultra} proposed a sleep staging ASIC that relies on wavelet packet decomposition (WPD) followed by handcrafted feature extraction and classification using a neural network-based decision tree. Their system, fabricated in 180nm CMOS, supports both EEG+EMG and EEG-only configurations. It achieves an accuracy of 81.1\% on NSRR and NTHU datasets, although the evaluation was not performed using standard k-fold validation, but instead relied on subset-based true/false positive analysis. The design consumes only 4.96\si{\micro\watt} and occupies 11.72\si{\milli\meter\squared}, showcasing extreme energy efficiency at the cost of generalizability and model flexibility.
}

% Another approach presented by Imtiaz \textit{et al.} is based around the extraction of 30 features from a single EEG input followed by a decision tree to sequentially test the most likely sleep stage if it is predicted to have changed from the previous epoch  \cite{imtiaz2017ultralow}. With this technique, they achieve 78.9 \% accuracy \cite{imtiaz2015automatic} using 0.58\si{\milli\watt} of power on a 10.3\si{\milli\square\meter} design.

\textcolor{black}{
Imtiaz \textit{et al.}~\cite{imtiaz2017ultralow, imtiaz2015automatic} developed a low-power system that extracts 30 features from a single EEG channel and uses a decision tree to classify sleep stages, conditional on detecting a transition from the previous epoch. Their implementation achieves an accuracy of 78.9\% while consuming 0.58\si{\milli\watt} on a 10.3\si{\milli\meter\squared} die. While efficient, the rule-based structure limits adaptability to new datasets or learning-based improvements.
}

% A final approach by Li \textit{et al.} employs a series of FPGA-controlled, time-multiplexed analog FIR filters to extra the $\theta$ and $\delta$ bands from 2 EEG and 1 EMG input channels \cite{li2016compact}. The device can only distinguish between REM, non-REM and wake sleep cycles. It does so by thresholding the \(\frac{\theta}{\delta}\) ratio. The design is based on a custom, general-purpose 64-channel neural network analyzer \cite{abdelhalim201364} controlled by an FPGA on a different chip. As such, it isn't as integrated as the other works but nonetheless presents an interesting approach to sleep staging. It must also be noted that the system is validated on a dataset consisting of data from 9 mice.

\textcolor{black}{
Another approach by Li \textit{et al.}\cite{li2016compact,kassiri2016electronic} employs FPGA-controlled, time-multiplexed analog FIR filters to extract the $\theta$ and $\delta$ frequency bands from two EEG and one EMG channel. Sleep stages are classified into wake, REM, and non-REM based on a threshold applied to the $\frac{\theta}{\delta}$ ratio. The system is part of a broader 64-channel neural signal analyzer platform\cite{abdelhalim201364}, controlled externally via an FPGA. While innovative, this approach is less integrated and only demonstrated on a dataset comprising recordings from nine mice, limiting its relevance to human applications. In contrast, the SleepViT architecture introduced in this work combines a modern attention-based model with a fully custom digital backend, offering both competitive accuracy and scalability for real-world, wearable deployment.
}

\section{Discussion}
\label{sec:discussion}
% The results presented in this work show that a small vision transformer can be implemented in hardware with a single EEG channel and achieve state-of-the-art accuracy. Indeed, Rong \textit{et al.} showed that most wearable devices for health applications have battery capacities around 100mAh \cite{rong2021energy}, which is far more energy than the accelerator requires over a full night if we consider an average power consumption of 0.56mW when including power gating. In addition, even in the relatively large 65nm CMOS process, the accelerator is small enough for any area budgets.

The results presented in this work demonstrate that a compact Vision Transformer model can be effectively implemented in hardware for real-time sleep stage classification using only a single EEG channel, achieving accuracy comparable to state-of-the-art methods. {\color{black}Although ViT implementations are typically memory-intensive, our design achieves a compact size along with the desired performance and power efficiency—without requiring sparsity-based optimizations. The achieved accuracy of 82.9\% is appropriate for the task of sleep staging. The American Academy of Sleep Medicine reported an inter-scorer agreement of only 82.6\% among more than 2,500 scorers, most of whom had over three years of experience~\cite{rosenberg2013american}. Moreover, the impact of misclassification in sleep staging is less critical compared to applications like seizure detection. In this context, prioritizing longer battery life, potentially compromised by larger models, is more important for improving user convenience.}

More importantly, this performance is achieved with exceptionally low power consumption. As shown by Rong \textit{et al.}~\cite{rong2021energy}, most wearable health-monitoring devices operate with battery capacities on the order of 100mAh \textcolor{black}{at 3.7V}. At an average power consumption of 0.56\si{\milli\watt}—including estimated leakage reductions from power gating—the proposed ASIC could operate continuously for multiple nights on a single charge. Furthermore, despite being fabricated in a relatively mature 65nm CMOS process, the total area of 0.754\si{\milli\meter\squared} is well within the area budget of modern wearable systems. \textcolor{black}{This work also demonstrates that competitive accuracy and power consumption can be achieved while relying solely on one electrode. This provides enhanced flexibility for practical embedded applications where a cumbersome multi-electrode system is undesirable, such as in-ear sleep staging for an acoustic neuromodulation treatment device.}

\section{Conclusion}
\label{sec:conclusion}
% This paper presents a hardware implementation of a lightweight vision transformer for sleep staging using a single EEG channel. The model, 8x times smaller than state-of-the-art, achieves an accuracy of 82.9\% on the MASS SS3 dataset. Its implementation in 65nm CMOS requires only 0.754\si{\milli\square\meter} of silicon, consumes an effective power of 0.56\si{\milli\watt} when considering power gating and has an inference latency of 45.6\si{\milli\second} and, as such, the device is suitable for a wearable device.

This work presents SleepViT, a hardware-accelerated Vision Transformer designed for low-power, real-time sleep stage classification using a single EEG channel. The proposed model achieves a classification accuracy of 82.9\% on the MASS SS3 dataset while using only 31.6k parameters, making it significantly smaller than other compact models reported in the literature.

The design is implemented in 65nm CMOS technology, occupying just 0.754\si{\milli\meter\squared} of silicon area. Each inference is completed in 45.6\si{\milli\second}, which corresponds to only 0.15\% of a 30-second sleep epoch—ensuring negligible latency. With module-level power gating applied, the effective average power consumption is reduced to 0.56\si{\milli\watt}, making the system suitable for always-on operation in wearable, battery-powered devices.

To the best of our knowledge, this is the first ASIC implementation of a Vision Transformer for sleep stage classification. The results demonstrate that transformer-based models, when co-designed with application-specific hardware optimizations such as fixed-point quantization and shared compute logic, can be effectively deployed in resource-constrained edge environments.

Future work will explore further power and memory optimizations, on-chip learning, and multimodal biosignal integration to extend the applicability of the platform to broader wearable healthcare use cases. \textcolor{black}{Sparse attention mechanisms or weight pruning techniques may be explored to further improve efficiency. Additionally, future work will assess how power, performance and area scale at newer technology nodes for this application, keeping in mind the target of a broad-market wearable device.}

\bibliographystyle{IEEEtran.bst}
\bibliography{IEEEabrv.bib,iscas_2025.bib}

\ifCLASSOPTIONcaptionsoff
  \newpage
\fi

\begin{comment}

\end{comment}

% that's all folks
\end{document}